\documentclass[letter,traditabstract]{aa}
\usepackage{txfonts}
\usepackage{graphicx}
\begin{document}


\title{{\it Herschel} observations of FIR emission lines in brightest cluster galaxies
\thanks{{\it Herschel} is an ESA space observatory with science instruments provided by European-led Principal Investigator consortia and with important participation from
NASA.}}

\author{A. C. Edge\inst{\ref{inst1}} \and J. B. R. Oonk\inst{\ref{inst2}} \and R. Mittal\inst{\ref{inst3}} \and 
S. W. Allen\inst{\ref{inst4}} \and
S. A. Baum\inst{\ref{inst3}} \and
H. B\"ohringer\inst{\ref{inst5}} \and
J. N. Bregman\inst{\ref{inst6}} \and
M. N. Bremer\inst{\ref{inst7}} \and
F. Combes\inst{\ref{inst8}} \and
C. S. Crawford\inst{\ref{inst9}} \and
M. Donahue\inst{\ref{inst10}} \and
E. Egami\inst{\ref{inst11}} \and
A. C. Fabian\inst{\ref{inst9}} \and
G. J. Ferland\inst{\ref{inst12}} \and
S. L. Hamer\inst{\ref{inst1}} \and
N. A. Hatch\inst{\ref{inst13}} \and
W. Jaffe\inst{\ref{inst2}} \and
R. M. Johnstone\inst{\ref{inst9}} \and
B. R. McNamara\inst{\ref{inst14}} \and
C. P. O'Dea\inst{\ref{inst15}} \and
P. Popesso\inst{\ref{inst5}} \and
A. C. Quillen\inst{\ref{inst16}} \and
P. Salom\'e\inst{\ref{inst8}} \and
C. L. Sarazin\inst{\ref{inst17}} \and
G. M. Voit\inst{\ref{inst10}} \and
R. J. Wilman\inst{\ref{inst18}} \and
M. W. Wise\inst{\ref{inst19}}
}

\institute{ Institute for Computational Cosmology, Department of Physics, Durham University, Durham, DH1 3LE, UK \label{inst1} \and
 Leiden Observatory, Leiden University, P.B. 9513, Leiden 2300 RA, The Netherlands \label{inst2} \and
 Chester F. Carlson Center for Imaging Science, Rochester Institute of Technology, Rochester, NY 14623, USA \label{inst3} \and
 Kavli Institute for Particle Astrophysics and Cosmology, Stanford University, 452 Lomita Mall, Stanford, CA 94305-4085, USA \label{inst4} \and
 Max-Planck-Institut f\"ur extraterrestrische Physik, 85748 Garching, Germany\label{inst5} \and
 University of Michigan, Dept. of Astronomy, Ann Arbor, MI 48109, USA \label{inst6} \and
 H H Wills Physics Laboratory, Tyndall Avenue, Bristol BS8 1TL, UK\label{inst7} \and
 Observatoire de Paris, LERMA, CNRS, 61 Av. de l'Observatoire, 75014 Paris, France\label{inst8} \and
 Institute of Astronomy, Madingley Rd., Cambridge, CB3 0HA, UK\label{inst9} \and
 Michigan State University, Physics and Astronomy Dept., East Lansing, MI 48824-2320, USA\label{inst10} \and
 Steward Observatory, University of Arizona, 933 N. Cherry Avenue, Tucson, AZ 85721, USA\label{inst11} \and
 Department of Physics, University of Kentucky, Lexington KY 40506 USA\label{inst12} \and
 School of Physics and Astronomy, University of Nottingham, University Park, Nottingham NG7 2RD, UK\label{inst13} \and
 Department of Physics \& Astronomy, University of Waterloo, 200 University Avenue West, Waterloo, Ontario, Canada N2L 3G1 \label{inst14} \and
 Department of Physics, Rochester Institute of Technology, 84 Lomb Memorial Drive, Rochester, NY 14623-5603, USA \label{inst15} \and
 Department of Physics and Astronomy, University of Rochester, Rochester, NY 14627, USA\label{inst16} \and
 Department of Astronomy, University of Virginia, P.O. Box 400325, Charlottesville, VA 22904-4325, USA\label{inst17} \and
 School of Physics, University of Melbourne, Victoria 3010, Australia\label{inst18} \and
 ASTRON, Netherlands Institute for Radio Astronomy,P.O. Box 2, 7990 AA Dwingeloo, The Netherlands\label{inst19}
}

\abstract{The question of how much gas cools in the cores of clusters 
of galaxies has been the focus of many, multiwavelength 
studies in the past 30 years.
In this letter we present the first detections of the 
strongest atomic cooling lines, [C{\sc ii}], [O{\sc i}] and [N{\sc ii}]  
in two strong cooling flow clusters, A1068 and A2597, using {\it Herschel} \rm PACS. 
These spectra indicate that the substantial
mass of cold molecular gas ($>10^{9}$M$_\odot$) known to be present in these
systems is being irradiated by intense UV radiation, most probably from young stars.
The line widths of these FIR lines indicate that they share dynamics similar
but not identical to other ionised and molecular gas traced by optical, near-infrared and CO
lines. 
The relative brightness of the FIR lines compared to
CO and FIR luminosity is consistent with other star-forming 
galaxies indicating that the properties of the molecular
gas clouds in cluster cores and the stars they form are not unusual.
These results provide additional evidence for a reservoir
of cold gas that is fed by the cooling of gas in the cores of the
most compact clusters and provide important diagnostics of
the temperature and density of the dense clouds this gas resides in.}

\date{Received 30 March 2010/Accepted}


\maketitle

\section{Introduction}

The cooling process at the cores of galaxy clusters is highly complex: recent {\it XMM-Newton} and {\it Chandra}
observations indicate that the cooling rates are
reduced by an order of magnitude below the simple cooling flow models at temperatures
below $\sim2\times10^7$K (Peterson \& Fabian 2006).
These X-ray observations, when linked with the detection
of radio jet inflated bubbles in the cores of many of the strongest cooling
flows (see McNamara \& Nulsen 2007 for a review), suggest that the
strong suppression of gas cooling is related to energy injection into the
intracluster medium by the action of jets and related AGN activity.

The detection of substantial masses of molecular gas in the cores
of the most rapidly cooling clusters through CO lines (Edge 2001, Salom\'e \& Combes 2003) and
warm H$_2$ molecular lines in the NIR and MIR (Jaffe \& Bremer 1997, Egami et al. 2006)
indicates that not all cooling is suppressed and this cooled gas may
provide the fuel for future AGN activity. These tracers of molecular gas appear
to correlate with the strength of optical lines from ionised gas (Crawford et al. 1999,
Edge 2001) and the dust continuum at MIR and sub-mm wavelengths (O'Dea et al. 2008).
However, the excitation of these various emission lines and the relative
importance of energy input from star formation, AGN, cosmic rays and/or the intracluster medium 
is poorly constrained (Ferland et al. 2009). 

One as yet unexplored diagnostic of the properties of the cold gas
are the atomic cooling lines found in the FIR, [C{\sc ii}], [O{\sc i}] and [N{\sc ii}].
The unprecedented sensitivity of {\it Herschel} \rm (Pilbratt et al. 2010)
to FIR line emission offers the
opportunity to assess the ionisation and density of the 
colder gas for the first time with the [C{\sc ii}] line 
and two principle [O{\sc i}] lines.
The authors were awarded 140~hours of time
in an Open Time Key Project (PI Edge) to investigate the FIR line and
continuum properties of a sample of 11 brightest cluster galaxies (BCGs)
in well-studied cooling flow clusters selected on the basis of optical
emission line and X-ray properties. 
The full goals of the 
project are to observe at least five atomic cooling lines for
each object that cover a range in density and temperature
behaviour and obtain a fully sampled FIR spectral energy
distribution for systems where significant star formation is expected. In this paper we present
the Photodetector Array Camera \& Spectrometer (PACS, Poglitsch et al. 2010) spectroscopy for the two targets 
observed in the Science Demonstration
Phase (SDP), Abell~1068 ($z=0.1386$) and Abell~2597 ($z=0.0821$). In
a parallel paper (Edge et al. 2010), we present the FIR photometry for
these clusters.

The two clusters observed have quite contrasting
 multiwavelength properties. A1068 is a 
strong MIR source (O'Dea et al 2008) with a 
bright CO detection (Edge 2001) but a weak radio source
(McNamara et al 2004). A1068 lies just below the luminosity 
threshold of a ULIRG (10$^{12}$~L$_\odot$) and exhibits
some contribution from an AGN (Crawford et al. 1999,
O'Dea et al. 2008). On the other hand, A2597 
is a relatively weak MIR source (Donahue et al. 2007)
with a weak CO detection (Salom\'e, priv. comm.) but
a powerful radio source (Sarazin et al. 1995). The implied
FIR luminosity of A2597  of  8.8$\times 10^{9}$~L$_\odot$ is a factor of around 30 below that
of A1068  (3.5$\times 10^{11}$~L$_\odot$) and,  in addition, the fractional contribution from an AGN 
in the MIR is also lower.

\section{Observations}

We have observed the [C{\sc ii}] and [O{\sc i}] lines at 157.74$\mu$m and 63.18$\mu$m for 
A1068 and A2597 with the PACS spectrometer on {\it Herschel}. These are the primary cooling lines
of the cold gas at a temperature T$<$40 K (Kaufman et al. 1999). In addition for A2597 we observed the 
[N{\sc ii}], [O{\sc i}] and [O{\sc iii}] lines at 121.90$\mu$m, 145.52$\mu$m and 88.36$\mu$m. These
lines are used to constrain the excitation and temperature of this gas. Table 1 gives a summary of the observations.

All spectral line observations were taken in \textit{PACS chopped line scan (standard faint line)} 
mode with chopping-nodding. The \textit{simple pointed observations} mode was used for
all observations.  The data were reduced following the PACS data reduction guide (PDRG) using the 
PACS Line Spectroscopy script for Point Source Chop/Nod Mode as
presented by the PACS ICC team during the {\it Herschel} science demonstration phase data processing 
workshop at ESAC in december 2009. The reduction was performed within
the {\it Herschel} Interactive Processing Environment (HIPE) version 2.0.0 (Ott 2010), build RC 3. We 
have processed the data from level 0 (raw channel data) to level 2
(calibrated spectra) in a number of steps as outlined in the PDRG.
Level 0 to 1.0 processing removes the telescope specific structures from the data. The slopes of 
the raw channel data are fitted and removed. The signal is
converted from data units to volts per second. Sky coordinate information is added and bad pixels 
and glitches are removed from data. The data is flatfielded
and flux calibrated by applying the ground based nominal response function as recommended in the 
PACS spectroscopy performance and calibration (PSPC) document. This
ground based response calibration is known to yield overestimated fluxes and following the PSPC 
we divide our fluxes by 1.3 and 1.1 in the blue and red bands.
The accuracy of this flux calibration for the PACS spectrometer, at the time of writing, is about 
50 percent within a given spectral band (PSPC).\par During the
final stage of the reduction, level 1.0 to 2.0, the data are spectrally and spatially rebinned 
into a 5$\times$5$\times$lambda cube. Using the standard 5$\times$5 spatial rebinning each
spatial pixel (spaxel) has a projected size of 9.4$''\times 9.4''$ on the sky. The spectral 
rebinning is performed using the recommended weak line density i.e. oversamp=1
and upsamp=4. Values between 1 and 10 were tried for the upsamp and 
oversamp parameters to test the robustness of the line profiles. We find
that the line profiles do not change significantly for this range in values.

\begin{table*}
\caption{Log of {\it Herschel}\rm PACS observations }
\begin{tabular}{lllccccc}
\hline\hline
Cluster & Redshift & Line  &  Wavelength & Obsid  & Bandwidth & Resolution  & Beam Size\\
        &          &       & ($\mu$m)   &        &(km~s$^{-1}$) & (km~s$^{-1}$) &  \\
\hline
A1068   & 0.1386   & [C{\sc ii}] & 179.61 & 1342186308 & 1200 & 201 & 13.5$''$/33~kpc  \\
        &          & [O{\sc i}]  &  71.94 & 1342186307 &  600 &  55 &  5.4$''$/13~kpc \\
A2597   & 0.0821   & [C{\sc ii}] & 170.78 & 1342187125 & 1100 & 218 & 12.8$''$/20~kpc \\
        &          & [O{\sc i}]  &  68.41 & 1342187124 &  550 &  68 &  5.1$''$/7.8~kpc\\
        &          & [N{\sc ii}] & 131.94 & 1342188942 & 1200 & 281 &  9.9$''$/15~kpc \\
        &          & [O{\sc i}b]  & 157.56 & 1342188704 & 1200 & 241 & 11.8$''$/18~kpc \\
        &          & [O{\sc iii}] & 95.61 & 1342188703 &      & 108 &  7.2$''$/11~kpc \\
\hline
 \end{tabular}
\end{table*}

\section{Results}

The [C{\sc ii}] 157$\mu$m and [O{\sc i}] 63$\mu$m lines are detected at a signal to noise greater 
than 30 for both A1068 and A2597. The much weaker 
[N{\sc ii}] 122$\mu$m and [O{\sc i}b] 145$\mu$m lines are
detected at the 3--5$\sigma$ level for A2597. The [O{\sc iii}] 88$\mu$m line was not detected 
in A2597, an upper limit for this line is given in Table 2.

The line spectra are fitted by a model consisting of; (i) a linear function to determine the 
continuum flux, and (ii) a single gaussian function to determine the
line flux. Continuum subtracted line spectra are shown for the central spaxel in Fig. 1 for 
A1068 and Fig. 2 for A2597. The fitted line centers agree well with the
redshift of CO in the BCG and the fitted FWHM line widths indicate gas with 
velocities of 300--500~km~s$^{-1}$. 

The [C{\sc ii}] and [N{\sc ii}] lines in the central spaxel of both objects are well 
described by a single gaussian. However, the [O{\sc i}] 63$\mu$m lines, 
where the PACS spectral resolution is best,
have profiles indicative of weak (2--3$\sigma$) deviations
from a single gaussian function. The [O{\sc i}]  line in A1068 hints at a two-component 
structure in the form of a narrow core component on top of a broad
underlying component  comparable to the CO(2-1) profile in Edge (2001). 
Both [O{\sc i}] lines observed in A2597 appear to have their dominant flux 
component at the systemic redshift of the BCG and a weaker component
offset by about +250~km~s$^{-1}$ which is also seen in the CO data (Salom\'e, priv. comm). 
We attribute the shared structure of these atomic and moledular lines to gas kinematics rather than 
self-absorption as the observed emission is from a large number of clouds that
have much narrower intrinsic line width.

The resolution of PACS at the observed wavelengths varies from about 5$''$ for the [O{\sc i}] 63$\mu$m 
line to about 14$''$ for the [C{\sc ii}] 157$\mu$m line. We have investigated line
emission in all 25 spaxels of the PACS FoV. In all cases the line flux is dominated by 
the central spaxel. Summing up the flux in all 25 spectra and comparing it to
the flux in the central spaxel shows no evidence of excess line flux as compared to 
what is expected from a point source. In order to properly recover the full beam
line fluxes we have applied point source corrections (appendix A of the PSPC document) 
to the central spaxel integrated line fluxes. The results are listed in Table
2. This spatial resolution matches the best sub-mm interferometry results for CO 
(Edge \& Frayer 2003; Salom\'e \& Combes 2004) which implies
that most of the emission is on scales $<5''$ so we believe our PACS line fluxes 
can be compared to literature values without large beam corrections.

\begin{table*}
\caption{Spectral line results for {\it Herschel}\rm PACS observations }
\begin{tabular}{lllcccc}
\hline \hline
Cluster & Redshift & Line & Integrated Line Flux & Velocity offset & measured FWHM  & instrinsic FWHM \\
        &          &      & ($10^{-18}$ W~m$^{-2}$) & (km~s$^{-1}$)  &  (km~s$^{-1}$) &  (km~s$^{-1}$)\\
\hline 
A1068   & 0.1386   & [C{\sc ii}] & 104.7$\pm$1.8 & +25$\pm$55 & 378$\pm$40 & 320$\pm$55 \\
        &          & [O{\sc i}]  &  64.8$\pm$0.2 & +25$\pm$50 & 356$\pm$40 & 352$\pm$55  \\
A2597   & 0.0821   & [C{\sc ii}] &  58.5$\pm$1.9 & +15$\pm$60 & 463$\pm$40 & 408$\pm$55 \\
        &          & [O{\sc i}]  &  54.7$\pm$0.2 & +40$\pm$55 & 411$\pm$40 & 405$\pm$55 \\
        &          & [N{\sc ii}] &   3.8$\pm$1.3 & +30$\pm$60 & 578$\pm$90 & 505$\pm$110 \\
        &          & [O{\sc i}b] &   3.3$\pm$1.3 & --57$\pm$65 & 484$\pm$90 & 420$\pm$110 \\
        &          & [O{\sc iii}] &  $<2.9$      &  -          & -          & -  \\
\hline
 \end{tabular}
\end{table*}

\begin{figure*}
\centerline{\includegraphics[width=5.65cm,angle=90]{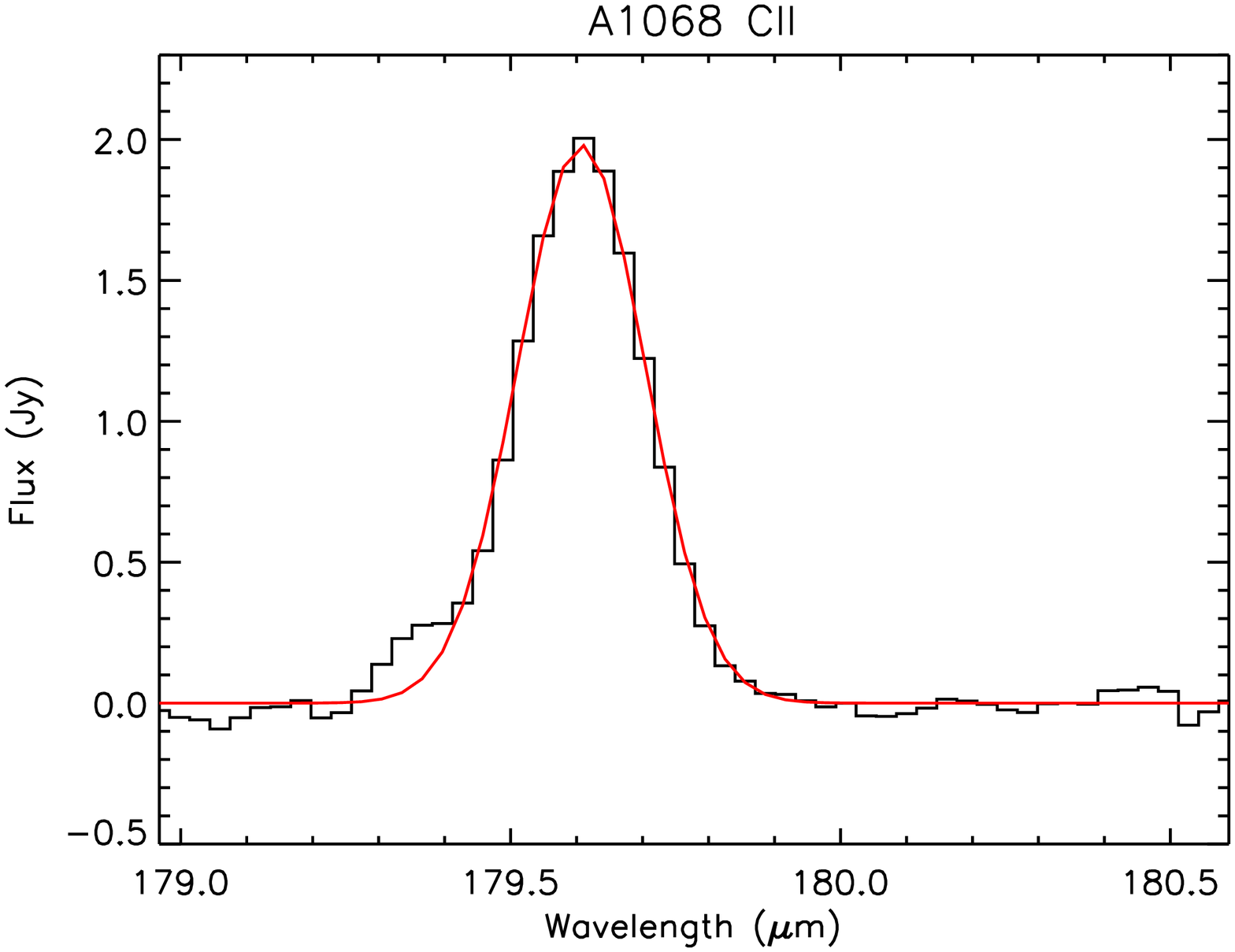}\includegraphics[width=5.65cm,angle=90]{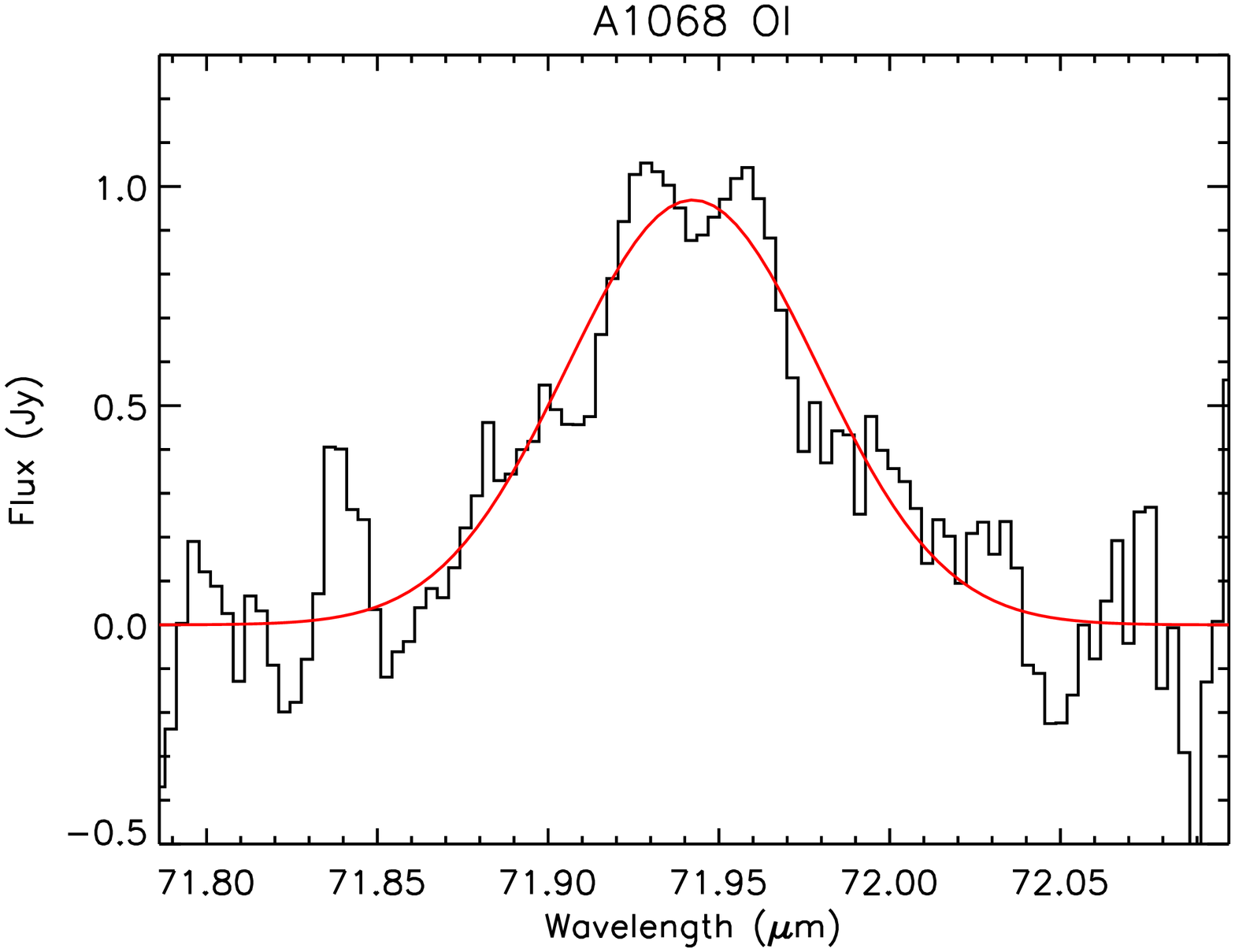}}
\caption{{\it Herschel} PACS spectra of [C{\sc ii}] and [O{\sc i}] in A1068}
\end{figure*}

\begin{figure*}
\centerline{\includegraphics[width=5.65cm,angle=90]{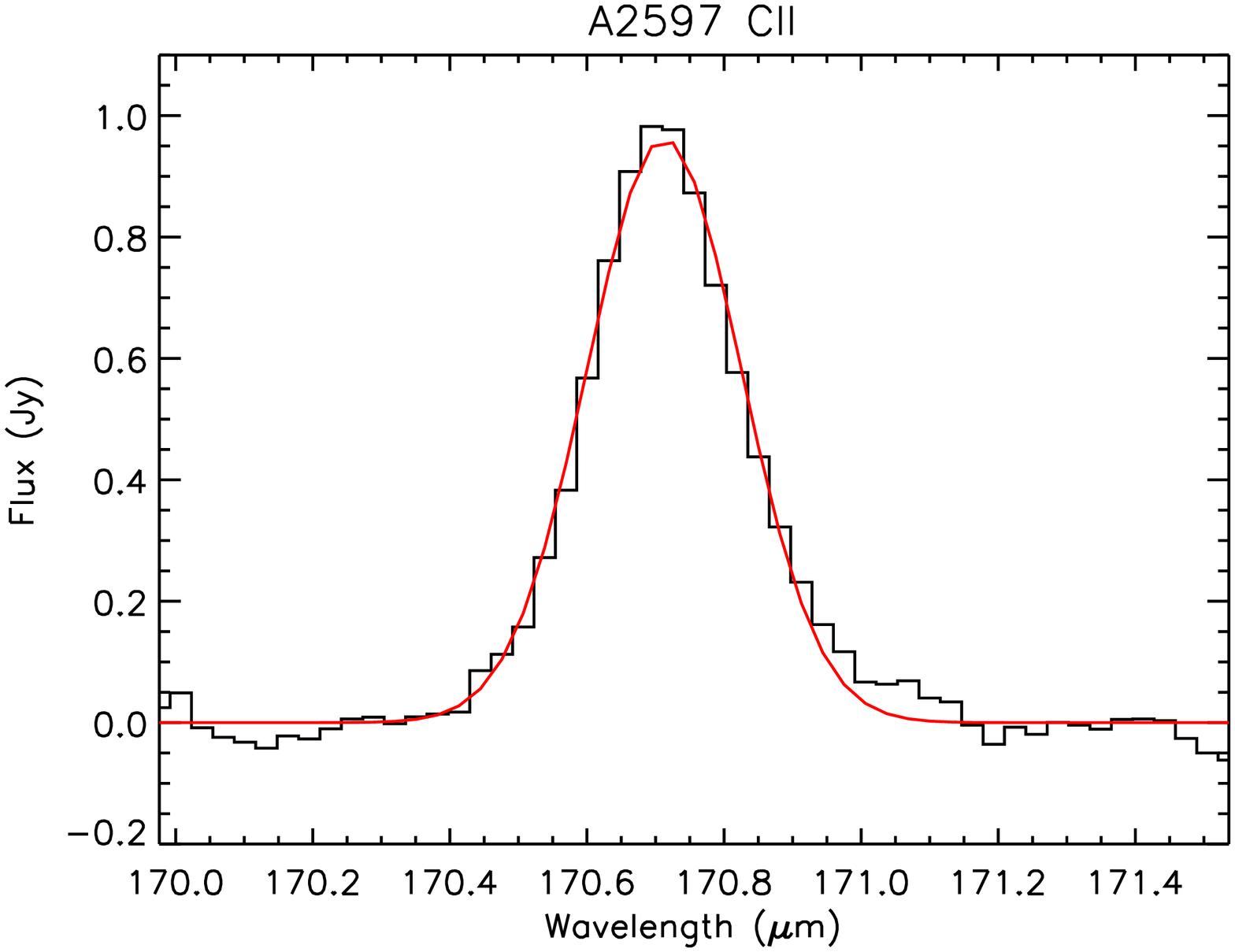}\includegraphics[width=5.65cm,angle=90]{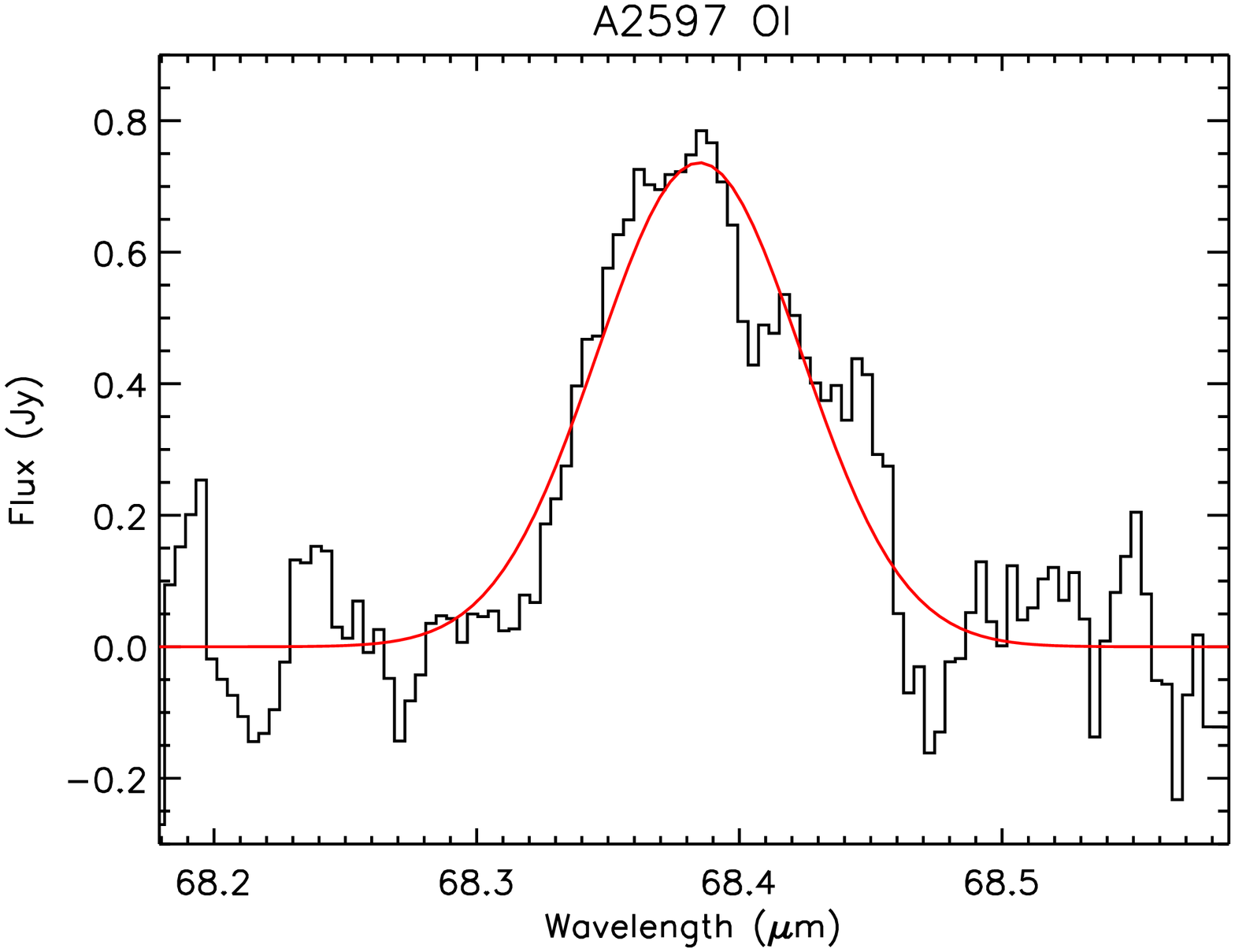}}
\centerline{\includegraphics[width=5.65cm,angle=90]{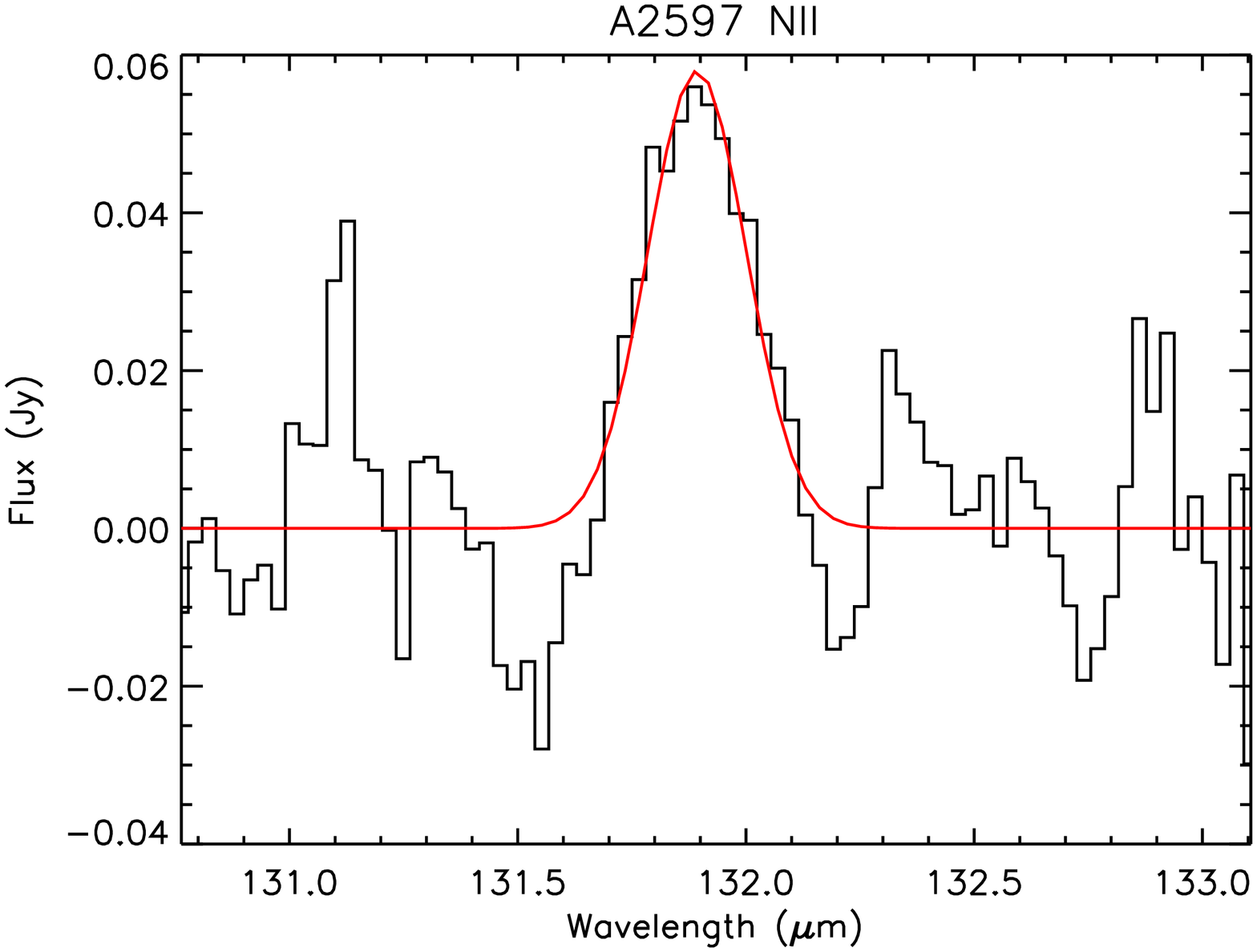}\includegraphics[width=5.65cm,angle=90]{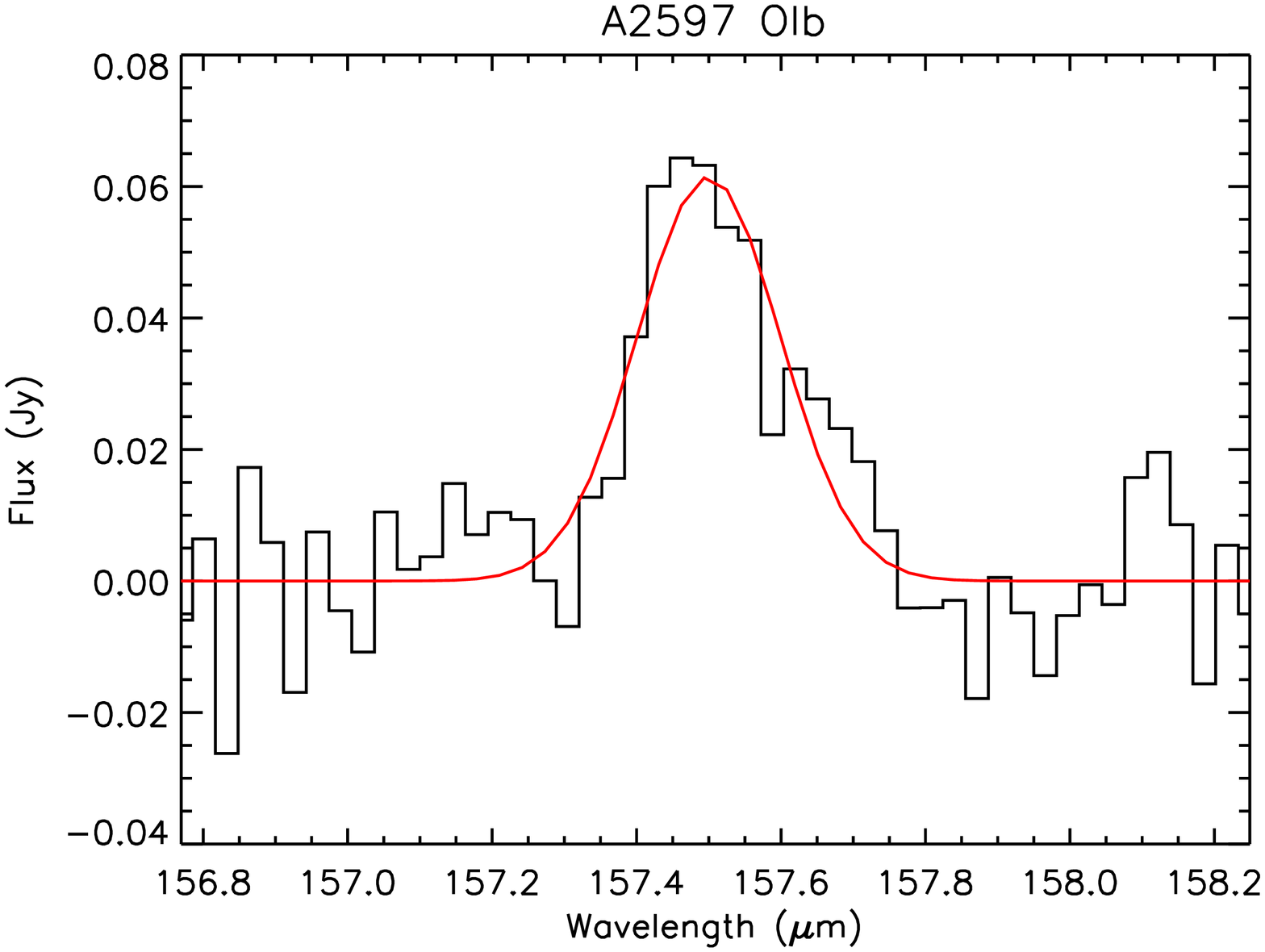}}
\caption{{\it Herschel} PACS spectra of [C{\sc ii}], [O{\sc i}] (63$\mu$m), [N{\sc ii}] and [O{\sc i}b] (145$\mu$m) in A2597.}
\end{figure*}

\section{Discussion}

The primary result from the SDP observations for
this project is that the atomic cooling lines are
present in both observed clusters. This first detection
of these lines in cluster cores reinforces the importance of
the cold gas in these environments. However, there
are a number of questions that these detections
raise.

{\it How do the properties of the FIR lines compare
to local LIRGs/ULIRGs?} \rm There have been several
studies of local galaxies with ISO  and high redshift galaxies 
using ground-based instrument that 
cover [C{\sc ii}] and [O{\sc i}] (Malhotra et al. 1997, Maiolino et al. 2005, Hailey-Dunsheath et al. 2010).
These studies show that the ratio of [C{\sc ii}] to
FIR luminosity is a function of luminosity with relatively
less  [C{\sc ii}] emission for the most FIR luminous 
sources. Using the  FIR data from Edge et al. (2010),
we calculate the [C{\sc ii}] to
FIR luminosity ratios are $10^{-2.4}$ and $10^{-1.9}$
for A1068 and A2597, respectively. 
The [C{\sc ii}]/FIR ratios of these two galaxies are comparable to 
those measured for galaxies of similar L$_{\rm FIR}$ (see Fig. 2 of
 Maiolino et al. (2005)).
In particular, the [C{\sc ii}] to FIR luminosity ratio is 
lower for the more FIR luminous of the two galaxies. The
ratio of  [C{\sc ii}] to  [O{\sc i}] shows less
variation (1.62 and 1.07) and is again consistent
with other comparable galaxies (Luhman et al. 2003).
Also our CO(1-0) to FIR luminosity ratios of
$10^{-5.7}$ and $10^{-5.6}$ for A1068 and A2597
are consistent with star-forming
local galaxies (Malhotra et al. 1997). So, despite potential
differences in excitation, pressure and metallicity, the
relative intensity of the atomic and molecular lines to
the FIR luminosity do not distinguish the BCGs studied here
from other FIR bright galaxies.

{\it How do the dynamics of atomic and molecular lines compare?} \rm
The relative velocity width of the
atomic lines compared to the CO and MIR H$_2$ lines
can provide important diagnostics for the dynamics
and energetics of the various gas tracers. From the
line width alone the resolution corrected line 
FWHM widths for the [C{\sc ii}]  and [O{\sc i}] lines are 
$\approx$330~km~s$^{-1}$ and $\approx$400~km~s$^{-1}$
for A1068 and A2597 respectively. This compares to
243$\pm$13~km~s$^{-1}$ (Edge 2001) and 292$\pm$45~km~s$^{-1}$
(Salom\'e, priv. comm.) for CO(2-1) for A1068
and A2597. In each case the FIR lines are 
a factor of $\sim$1.35 broader. This is not due to
any instrumental broadening in the PACS instrument
as the two lines sampled have similar intrinsic
line width despite being observed at very different
resolution. 
Instead, this difference is more likely to
be related to the lines being emitted from 
different regions within the BCG or in shocks.
However, this
clearly needs to be tested in more systems and
through direct comparison of the [C{\sc ii}] and [O{\sc i}]
extent with that of CO.

{\it How do the FIR line ratios constrain the gas properties?} \rm
The relative strength of the FIR lines can constrain
several key properties of the gas phase that dominates
the emission. The main constraint we can determine
directly from our current data is from the  [C{\sc ii}] 158$\mu$m and [O{\sc i}]
63 and 145$\mu$m lines for A2597.
Kaufman et al. (1999) present
photodissociation region (PDR) model predictions
for the [O{\sc i}] 145$\mu$m/63$\mu$m and [C{\sc ii}] 158$\mu$m to  [O{\sc i}]
63$\mu$m line ratios. Combining these two constraints for our
observed [O{\sc i}] 145$\mu$m/63$\mu$m ratio of 0.06$\pm$0.02 
and  [O{\sc i}] 63$\mu$m  to [C{\sc ii}] 158$\mu$m ratio of 0.94$\pm$0.05, we
estimate a density of 10$^{3.3\pm 0.5}$cm$^{-3}$ and an
incident FUV flux of G$_0$ of 150--1000 Habing units.
These values of G$_0$ imply intrinsic FUV luminosities of $\approx 2-5 \times 10^{43}$~erg~s$^{-1}$
if the clouds subtend 3--5~kpc. This is comparable to the observed FUV 
luminosities of these galaxies once dust absorption is taken into account (O'Dea et al. 2004).
 
\section{Conclusions}

These initial results from {\it Herschel} \rm
indicate that atomic cooling lines
are present in the
brightest cluster galaxies in cooling flow clusters.
The intensity and velocity width of these lines
is consistent with all the other observed tracers 
of cold gas in these systems implying they
originate from the same population of
clouds. The only apparent exception to this
in our current observations is that the 
FIR lines appear to be systematically 
broader than the CO lines impling that
the {\it relative} intensity of these lines
varies with position within the BCG. 
The results that will come from our Open Time
Key Project for 11 BCGs will expand greatly
on those presented here with more lines and 
a greater range of BCG properties. Beyond
this, the potential for {\it Herschel} \rm
to illuminate the properties of the cold
gas that may fuel cold nuclear accretion
in more distant clusters and local 
groups is vast.

\begin{acknowledgements}
We would like to thank the {\it Herschel} Observatory 
and instrument teams
for the extraordinary dedication they have shown to
deliver such a powerful telescope. We would
like to thank the HSC and NHSC consortium for help
with data reduction pipelines. J.B.R.O. thanks HSC, 
the {\it Herschel} Helpdesk and the PACS group at MPE for useful discussions.
R.\,M. thanks the NHSC for the HIPE tutorials.
\end{acknowledgements}

\end{document}